\documentclass[3p,twocolumn,times,number]{elsarticle}
\pdfoutput=1 
\usepackage{graphicx}
\usepackage{amsmath}   
\usepackage[left]{lineno}   
\usepackage{hyperref}

\newcommand{\bbonu}{\ensuremath{\beta\beta0\nu}}
\newcommand{\bbtnu}{\ensuremath{\beta\beta2\nu}}

\newcommand{\bb}{\ensuremath{\beta\beta}}

\begin{document}

\begin{frontmatter}

\title{The NEXT experiment: A high pressure xenon gas TPC for neutrinoless double beta decay searches}

\author[add1]{D. Lorca\corref{cor}}
\ead{david.lorca@ific.uv.es}
\author[add1]{J. Mart\'in-Albo}
\author[add1]{F. Monrabal}
\author{on behalf of the NEXT Collaboration}

\cortext[cor]{Corresponding author}

\address[add1]{Instituto de Fisica Corpuscular (IFIC), CSIC \& Univ. de Valencia, E-46071 Valencia, Spain}

\begin{abstract}

Neutrinoless double beta decay (\bbonu) is a hypothetical, very slow nuclear transition in which two neutrons undergo beta decay simultaneously and without the emission of neutrinos. The importance of this process goes beyond its intrinsic interest: an unambiguous observation would establish a Majorana nature for the neutrino and prove the violation of lepton number.

NEXT is a new experiment to search for neutrinoless double beta decay using a radiopure high-pressure xenon gas TPC, filled with 100 kg of Xe enriched in Xe-136.  NEXT will be the first large high-pressure gas TPC to use electroluminescence readout with SOFT (Separated, Optimized FuncTions) technology. The design consists in asymmetric TPC, with photomultipliers behind a transparent cathode and position-sensitive light pixels behind the anode.
The experiment is approved to start data taking at the Laboratorio Subterr\'aneo de Canfranc (LSC), Spain, in 2014. 
\end{abstract}


\end{frontmatter}

\section{Introduction}

Double beta decay (\bb) is a very rare nuclear transition in which a nucleus with Z protons decays into a nucleus with $Z+2$ protons and same mass number A. It can only be observed in those isotopes where the decay to the $Z+1$ isobar is forbidden or highly suppressed. Two decay modes are usually considered:

\begin{enumerate}
\item The standard two-neutrino mode (\bbtnu), consisting in two simultaneous beta decays, $^A _Z X \rightarrow{} ^A _{Z+2} Y + 2 e^- + 2 \bar{\nu_e}$, which has been observed in several isotopes with typical half-lives in the range of $10^{18} - 10^{21}$ years \cite{GomezCadenas:2011it}.

\item The neutrinoless mode (\bbonu\ ), $^A _Z X \rightarrow{} ^A _{Z+2} Y + 2 e^-$, which violates lepton-number conservation, and is therefore forbidden in the Standard Model of particle physics. An observation of \bbonu\ would prove that neutrinos are Majorana particles, that is identical to their antiparticles \cite{Schechter:1981bd}.
\end{enumerate}

The observation of \bbonu\ would demonstrate that total lepton number is violated, an observation that could be linked to the cosmic asymmetry between matter and antimatter through the process known as leptogenesis \cite{Davidson:2008bu}. Besides, a Majorana nature for the neutrino, may explain the smallness of neutrino masses \cite{GomezCadenas:2011it}. 

The half-life of \bbonu, if mediated by light, Majorana neutrino exchange, can be written as 

\begin{equation}
(T^{0\nu} _{1/2})^{-1} = G^{0\nu}  \left |M^{0\nu}\right |^2 m^2 _{\bb}
\end{equation}

where $G^{0\nu}$ is an exactly-calculable phase-space integral for the emission of two electrons; $\left |M^{0\nu}\right |$ is the nuclear matrix element of the transition, that has to be evaluated theoretically; and $m _{\bb}$ is the effective Majorana mass of the electron neutrino:

\begin{equation}
m _{\bb} = \left | \sum U^2 _{ei}  m_i\right |
\end{equation}

where $m_i$ are the neutrino mass eigenstates and $U_{ei}$ are elements of the neutrino mixing matrix.
Therefore a measurement of the \bbonu\ decay rate would provide direct information on neutrino masses \cite{GomezCadenas:2011it}.

\section{The NEXT concept}
Double beta decay experiments are designed to measure the energy of the radiation emitted by a \bb\ source. In the case of  \bbonu, the sum of the kinetic energies of the two emitted electrons is always the same, and corresponds to the mass difference between the parent and the daughter nuclei: $Q_{\bb\ }= M(Z,A) - M(Z+2,A)$. However, due to the finite energy resolution of any detector, \bbonu\  events are reconstructed within a non-zero energy range centered around $Q_{\beta\beta}$ , typically following a gaussian distribution. Other processes occurring in the detector can fall in that region of energies, thus becoming a background and compromising drastically the experiment's expected sensitivity to $m_{\beta\beta}$ \cite{GomezCadenas:2010gs}.

The NEXT experiment combines good energy resolution, a low background rate and the possibility to scale-up the detector to large masses of \bb\ isotope by using a high-pressure xenon gas (HPXe) electroluminescent time projection chamber (TPC) to search for \bbonu\ in Xe-136. The combination results in excellent sensitivity to $m_{\beta\beta}$ . For a total exposure of $500\ {\rm kg}\cdot{} {\rm year}$, the sensitivity is better than 100 meV. This sensitivity can match or outperform that of the best experiments in the field \cite{GomezCadenas:2010gs}.

Xenon is a suitable detection medium that provides both scintillation and ionization signals. In its gaseous phase, xenon can provide high energy resolution, better than 0.5\% at 2500 keV \cite{KlapdorKleingrothaus:2000sn}. The two-neutrino decay mode of Xe-136 is slow, $\sim 2.3 \cdot{}  10^{21}$ years \cite{Ackerman:2011gz}, and hence the experimental requirement for good energy resolution is less stringent than for other \bb\ sources. Xe-136 constitutes 8.86\% of all natural xenon, but the enrichment process is relatively simple and cheap compared to that of other \bb\ isotopes, thus making Xe-136 the most obvious candidate for a future multi-ton experiment. Also, xenon, unlike other \bb\  sources, has no long-lived radioactive isotopes that could become a background. 

\begin{figure}[tbhp!]
\begin{center}
\includegraphics[width=0.4\textwidth]{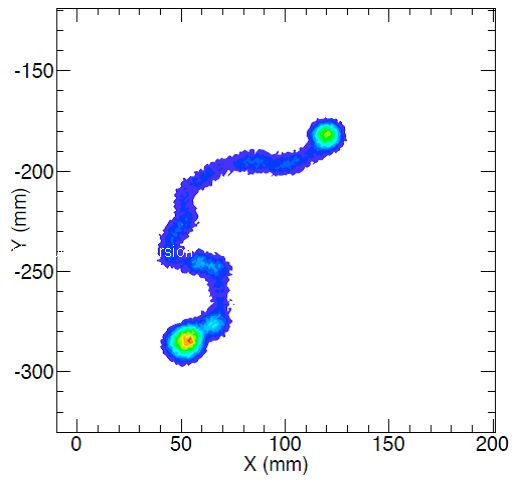} 
\end{center}
\caption{\small Monte-Carlo simulation of a Xe -136 \bbonu\ event in xenon gas at 10 bar: the ionization track, about 30 cm long, is tortuous because of multiple scattering, and has larger depositions or blobs in both ends.} 
\label{fig:track} 
\end{figure}

Neutrinoless double beta decay events leave a distinctive topological signature in gaseous xenon: an ionization track, about 30 cm long at 10 bar, tortuous due to multiple scattering, and with larger energy depositions at both ends, as shown in Figure \ref{fig:track}. This signature can be used for background identification and rejection.

To achieve optimal energy resolution, the ionization signal is amplified in NEXT using the electroluminescence light (EL) of xenon. The chamber will have separated detection systems for tracking and calorimetry. This is the so-called SOFT concept \cite{Nygren:2009zz, Granena:2009it}.

The detection process is as follows: Particles interacting in the HPXe transfer their energy to the medium through ionization and excitation. The excitation energy is manifested in the prompt emission of VUV $(\sim178\ {\rm nm})$ scintillation light. The ionization tracks left behind by the particle are prevented from recombination by an electric field. The ionization electrons drift toward the TPC anode, entering a region, defined by two highly-transparent meshes, with an even more intense electric field. There, further VUV photons are generated isotropically by electroluminescence. Therefore both scintillation and ionization produce an optical signal, to be detected with a plane of PMTs (the energy plane). The detection of the primary scintillation light constitutes the start-of-event, whereas the detection of EL light provides an energy measurement. Electroluminescent light provides tracking as well, since it is detected also a few millimeters away from production at the anode plane, via an array of MPPCs (the tracking plane).

\begin{figure}[tbhp!]
\begin{center}
\includegraphics[width=0.45\textwidth]{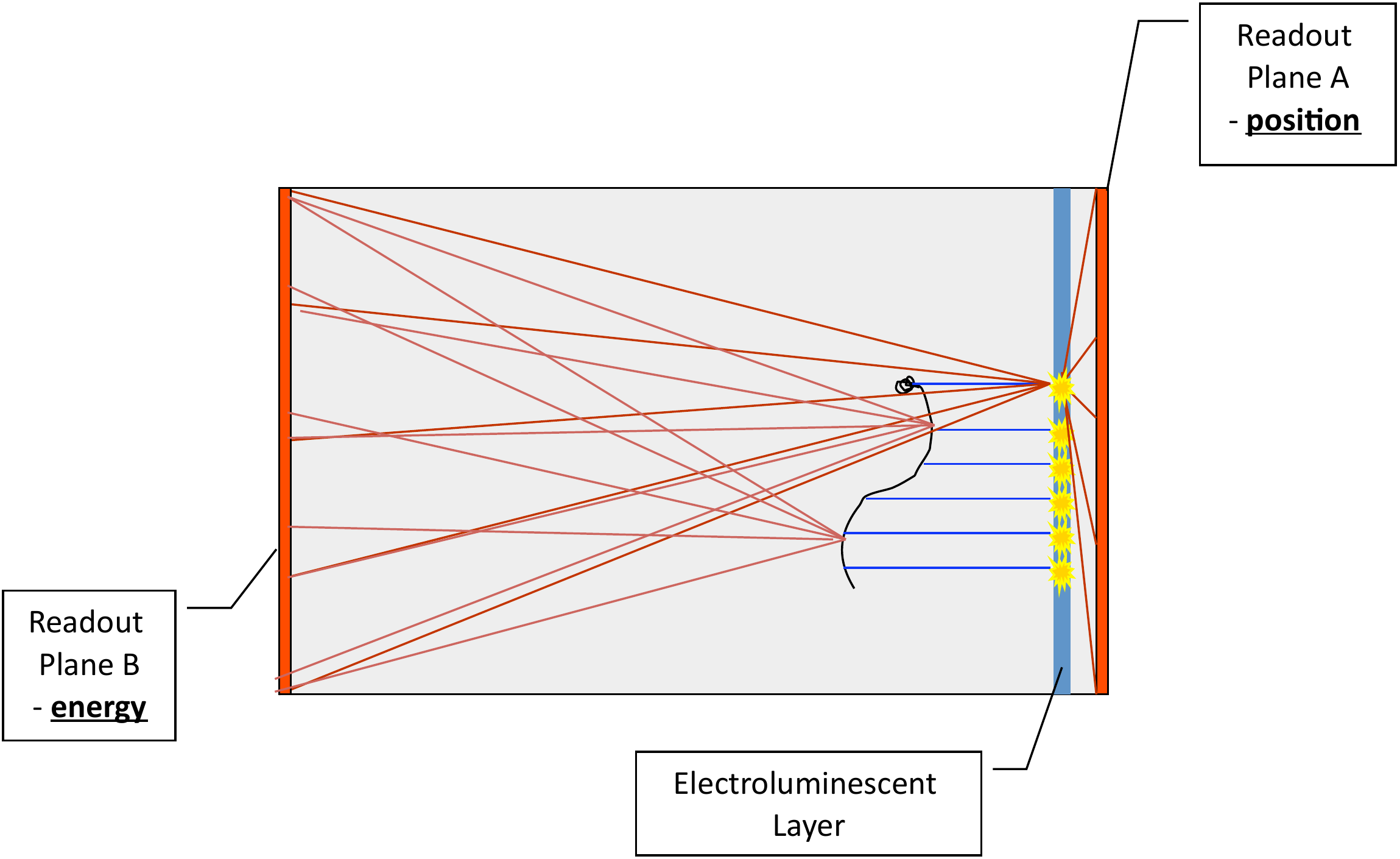} 
\end{center}
\caption{\small The SOFT concept. EL light generated at the anode is recorded in the photosensor plane right behind it and used for tracking. It is also recorded in the photosensor plane behind the transparent cathode and used for a precise energy measurement.}
\label{fig:TPC} 
\end{figure}

\section{The NEXT-DEMO Prototype} \label{NEXT-DEMO}

To demonstrate that the detector concept described in the previous section is reliable, a scaled prototype, NEXT-DEMO, was developed. This detector, operating at IFIC (Valencia, Spain), is equipped with an energy plane made of 19 Hamamatsu R7378A PMTs, and a tracking plane made of 248, $ 1\ {\rm mm^2}$, Hamamatsu MPPCs. The stainless-steel pressure vessel, 60 cm long and 30 cm diameter, can withstand 15 bar and thus holds about 10 kg of xenon \cite{Collaboration:2012ha}. The goals of the prototype are:

\begin{enumerate}
\item to demonstrate track reconstruction and the performance of MPPCs.
\item to test long drift lengths (30 cm) and high voltages (up to 50 kV in the cathode and 25 kV in the anode).
\item to understand gas recirculation in a large volume, including operation stability and robustness against leaks.
\item to understand the transmittance of the light tube, with and without wavelength shifter.
\item to demonstrate that the target energy resolution is realistic in a large-scale detector.
\end{enumerate}

After months of operation, the system has proven to be stable and millions of event have been acquired. Gas purity has been studied using RGA (mass spectrometer), and using a hot getter we have reached electron lifetimes of $\sim$ 3.6 ms. 

A reconfigurable and efficient hardware trigger system has been developed, allowing on-line triggering based on the detection of primary or secondary scintillation light, or a combination of both, of the selected PMT sensors.

\begin{figure}[tb]
\begin{center}
\includegraphics[width=0.45\textwidth]{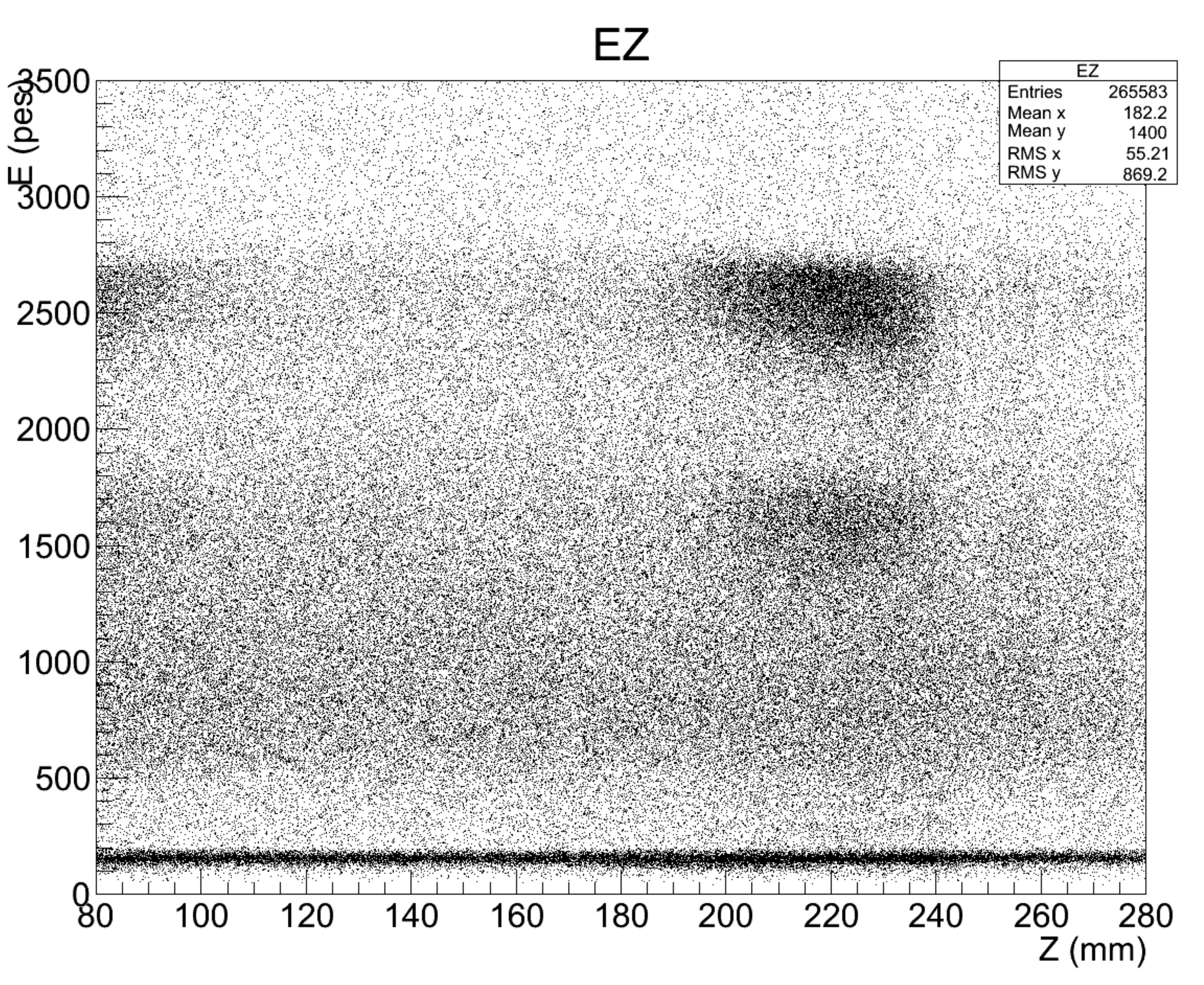} 
\includegraphics[width=0.45\textwidth]{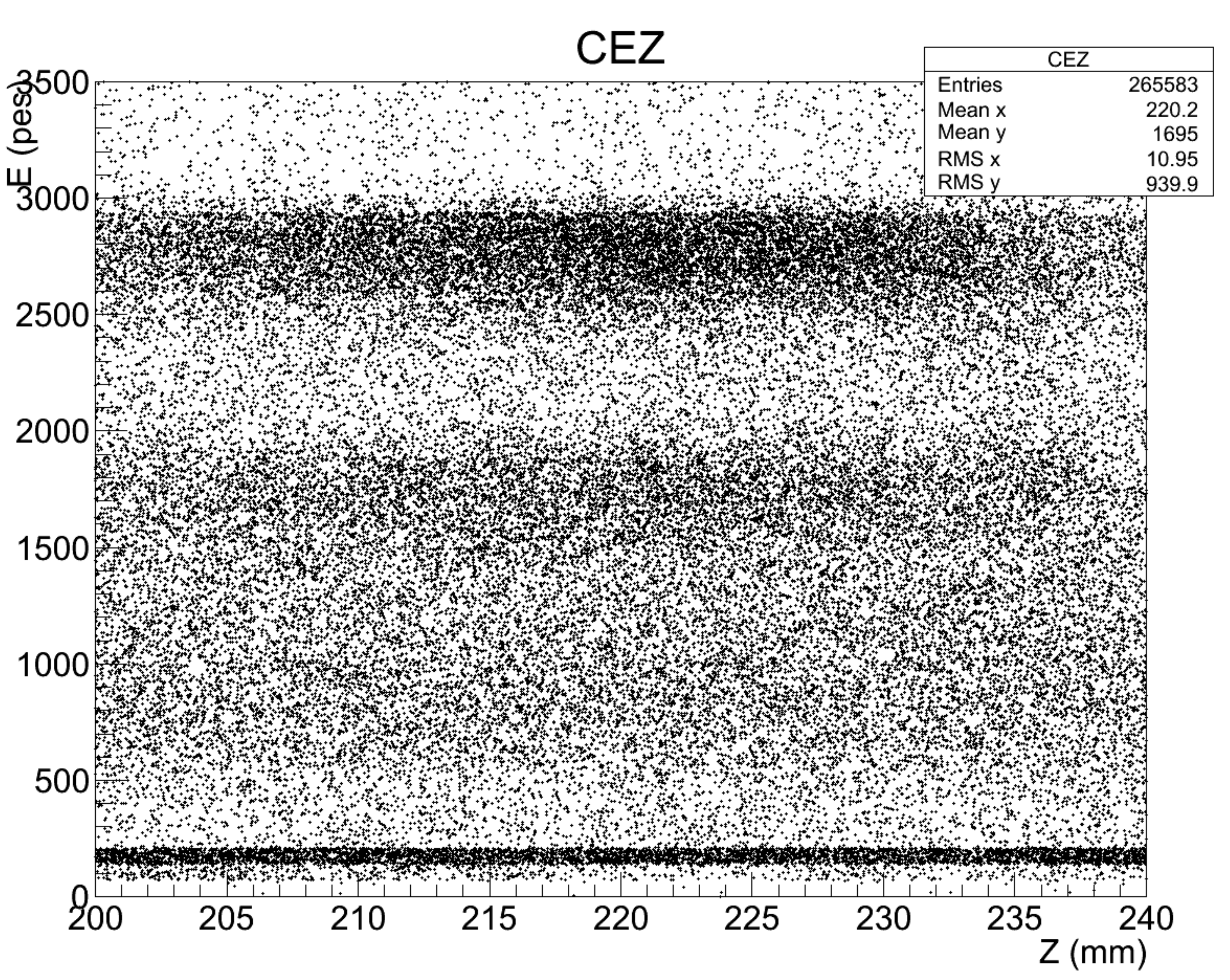} 
\end{center}
\caption{\small Top: Number of photolectrons detected decrease with distance in Z due to attachment of $e^-$ in gas. Bottom: Correction in Z axis.}
\label{fig:Zcorrection} 
\end{figure}

On the other hand, different algorithms have been developed for the correctly calibration of the TPC, correcting both spatial and temporal effects (see Figure \ref{fig:Zcorrection}).

\begin{figure}[tb]
\begin{center}
\includegraphics[width=0.45\textwidth]{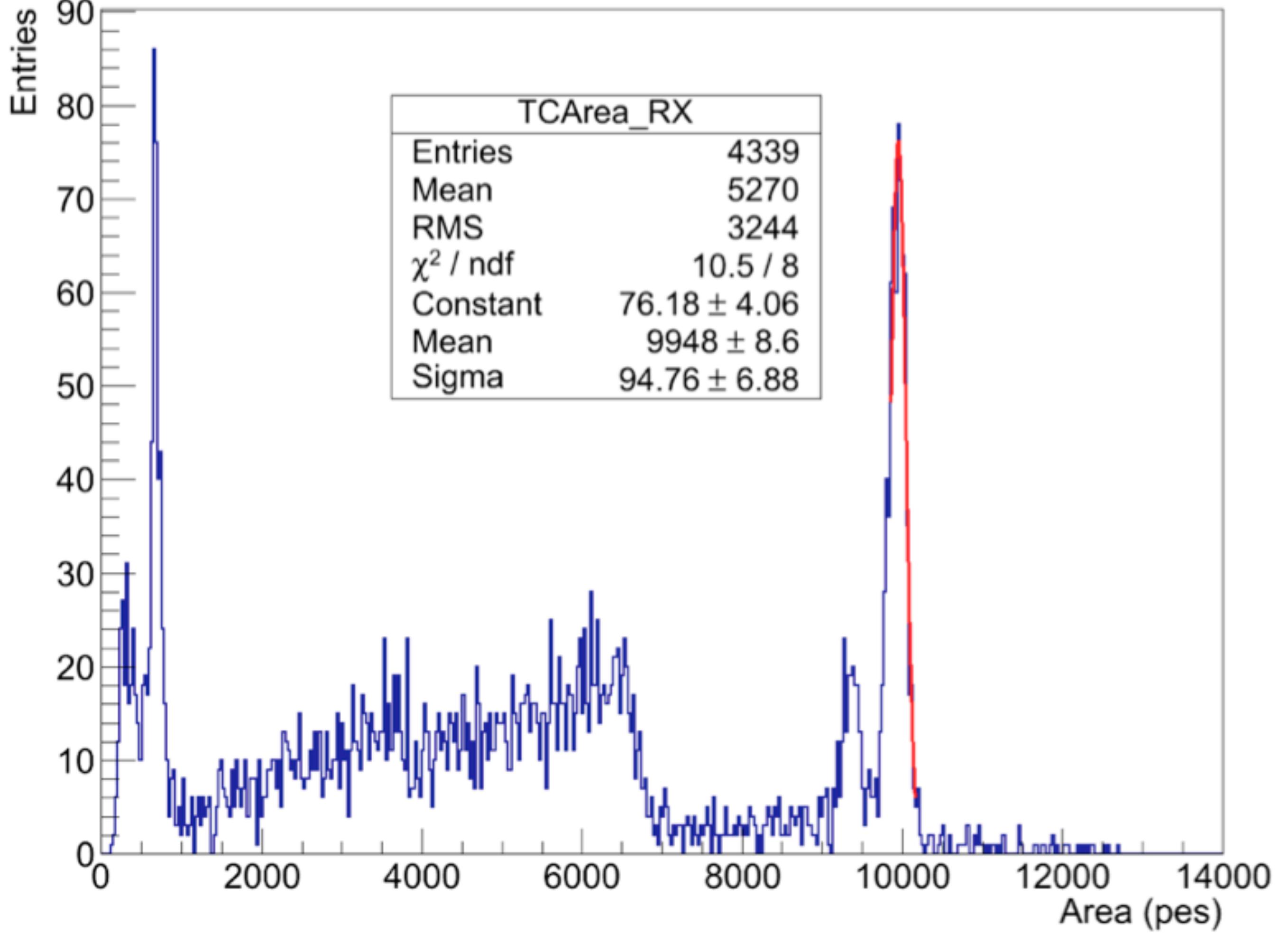} 
\end{center}
\caption{\small Spectrum of Na22 source, showing energy resolution of 2.24\% FWHM, which extrapolates to 1.02\% at $Q_{\bb}$.}
\label{fig:Spectrum} 
\end{figure}

At the same time, preliminary data analysis has already proven target energy resolution using a Na22 source (see Figure \ref{fig:Spectrum}). Tracking Plane with MPPCs is already being installed, and studies on topology reconstruction are underway.

\section{NEXT-100} \label{NEXT-100}

The NEXT Collaboration has recently published the Technical Design Report \cite{Collaboration:2012ha} of the NEXT-100 detector, result of the intense R\&D carried on during the previous three years. The design consists of a cylindrical  pressure vessel built with low activity stainless steel. To shield this activity we introduce an inner copper shield, 12\ {\rm cm} thick and made of radiopure copper, with an activity of about 5-10 $\mu{\rm Bq}/{\rm kg}$. This copper shield will attenuate the radiation coming from  the high-energy gammas emitted in the decays of $^{208}$Tl and $^{214}$Bi, present in the external detector, by a factor of 100 \cite{Collaboration:2012ha}.

Inside the pressure vessel, the field cage, a cylindrical shell 2.5 cm thick, will have three wire meshes (cathode, gate and anode), that define two electric field regions. The drift region, between cathode and gate, will be 107 cm diameter and 130 cm length, and will have a field strength of $0.3\ {\rm kV}\cdot{}{\rm cm^{-1}}$ . The EL region, between gate and anode, will be 0.5 cm long, and will have a (E/p) of $3.0\ {\rm kV}\cdot{}{\rm cm^{-1}}\cdot{}{\rm bar ^{-1}}$ , which produces around $2500\ {\rm photons/e^-}$ \cite{Collaboration:2012ha}.

At 0.5 cm away from the EL region, there is the Tracking Plane, in charge of the reconstruction of the track's events happening in the TPC. The MPPC chosen for NEXT-100 is the S10362-11-050P model by Hamamatsu. This device has an active area of $1\ {\rm mm^2}$, 400 sensitive cells ($50\ {\rm \mu m}$ size) and high photon detection efficiency (PDE) in the blue region (about $\sim 50\%$ at 440 nm). This PDE peaks in the blue region of the spectrum, and they are not sensitive below 200 nm, where the emission spectrum of xenon lies. Consequently, the sensors will be coated with tetraphenyl butadiene (TPB). The NEXT Collaboration has developed a procedure to deposit thin layers of TPB by vacuum evaporation \cite{Alvarez:2012ub}.

The energy measurement in NEXT-100 is provided by a total of 60 Hamamatsu R11410-10 photomultipliers covering 32.5\% of the cathode area constitute the energy plane. This phototube model has been specially developed for radiopure, xenon-based detectors. In NEXT-100 they will be sealed into individual pressure resistant, vacuum tight copper enclosures coupled to sapphire windows \cite{Collaboration:2012ha}.

To improve the light collection efficiency of the detector, reflector panels coated with the same wavelength shifter as MPPCs (TPB), will cover the inner part of the field cage.

\section{Conclusions} \label{Conclusions}

The search for \bbonu\ is one of the major current challenges in neutrino physics. Due to its high sensitivity to $m_{\bb}$, NEXT-100 promises to be one of the leaders in the field. The detector will use electroluminescence for the amplification of the ionization signal, and will have separate readout planes for tracking (an array of MPPCs) and calorimetry (an array of PMTs). Such a design provides both optimal energy resolution and event topological information for pattern recognition. This, together with the extreme radiopurity of the employed materials allow a background suppression of more than 7 orders of magnitude, giving an expected background rate of $ \simeq 8 \times 10^{-4}\ {\rm counts} /{\rm keV}\cdot{}{\rm kg}\cdot{}{\rm y}$ \cite{Collaboration:2012ha}.

 After several years of intense technological development, prototypes are already providing useful results for understanding the correct functioning of the final detector.

NEXT-100 is approved for operation in the Laboratorio Subterr\'aneo de Canfranc (LSC), in Spain. The installation of shielding and ancillary systems will start in the second half of 2012. The assembly and commissioning of the detector is planned for early 2014.

\section*{Acknowledgments}

This work was supported by the Spanish Ministerio de Econom\'ia y Competitividad under grants CONSOLIDER-Ingenio 2010 CSD2008-0037 (CUP) and FPA2009-13697-C04-04.

\nolinenumbers

\end{document}